\documentstyle[11pt,bezier]{article}
\begin{document}
\def \inbar{\vrule height1.5ex width.4pt depth0pt}
\def \xC{\relax\hbox{\kern.25em$\inbar\kern-.3em{\rm C}$}}
\def \xR{\relax{\rm I\kern-.18em R}}
\newcommand{\xZ}{Z \hspace{-.08in}Z}
\newcommand{\xbe}{\begin{equation}}
\newcommand{\xee}{\end{equation}}
\newcommand{\xbea}{\begin{eqnarray}}
\newcommand{\xeea}{\end{eqnarray}}
\newcommand{\xnn}{\nonumber}
\newcommand{\xkt}{\rangle}
\newcommand{\xbr}{\langle}
\newcommand{\xlll}{\left( }
\newcommand{\xrrr}{\right)}
\newcommand{\xcun}{\mbox{\footnotesize${\cal N}$}}
\title{On a Class of Quantum Canonical  Transformations and the
Time-Dependent Harmonic Oscillator}
\author{Ali Mostafazadeh\thanks{E-mail: alimos@phys.ualberta.ca}\\ \\
Theoretical Physics Institute, University of Alberta, \\
Edmonton, Alberta,  Canada T6G 2J1.}
\date{December 1996	}
\maketitle

\begin{abstract} 
Quantum canonical transformations corresponding to the action of the
unitary operator  $e^{i\epsilon(t)\sqrt{f(x)}p\sqrt{f(x)}}$ is studied.
It is shown that for $f(x)=x$, the effect of this transformation is to rescale the
position and momentum operators by  $e^{\epsilon(t)}$ and $e^{-\epsilon(t)}$,
respectively. This transformation is shown to lead to the identification of a
previously unknown class of exactly solvable time-dependent harmonic
oscillators. It turns out that the Caldirola-Kanai oscillator whose mass is
given by $m=m_0 e^{\gamma t}$, belongs to this class. It is also shown
that for arbitrary $f(x)$, this canonical transformations map the dynamics of
a free particle with constant mass to that of free particle with a
position-dependent mass. In other words, they lead to a change of the metric
of the space.
\end{abstract}

It is well-known that  in quantum mechanics the unitary transformations of the
Hilbert space correspond to the canonical transformations of the classical
mechanics. Unfortunately, these {\em quantum canonical transformations}
have not been usually discussed in most textbooks on quantum mechanics.
The purpose of this note is to study a class of canonical transformations, namely
	\xbe
	{\cal U}:=\exp\left[ \frac{i\epsilon(t)}{2}\{f(x),p\}\right]
	=\exp \left[i\epsilon(t)\sqrt{f(x)}\,p\,\sqrt{f(x)}\right]\;,
	\label{u}
	\xee
and demonstrate their utility in solving the Schr\"odinger equation for a class
of time-dependent harmonic oscillators. In Eq.~(\ref{u}), $x$ and $p$ denote
position and momentum operators, respectively, $\{~,~\}$ stands for the
anticommutator of two operators, and $f$ and $\epsilon$ are arbitrary functions.

Let us first recall the effect of a general time-dependent quantum canonical
transformation ${\cal U}={\cal U}(t)$ on the Hamiltonian $H=H(t)$ and the
time-evolution operator $U=U(t)$, i.e., the relations
	\xbea
	H(t)\rightarrow H'(t)&=&
	{\cal U}(t)H(t)\,{\cal U}^\dagger(t)-i\,{\cal U}(t)\,\dot{\cal U}^\dagger(t)\;,
	\label{trans}\\
	U(t)\rightarrow U'(t)&=&{\cal U}(t)U(t)\,{\cal U}^\dagger(0)\;,
	\label{trans-u}
	\xeea
where a dot means a time-derivative and $\hbar$ is set to unity.
These equations are direct  consequences of the requirement that the
Schr\"odinger equation
	\xbe
	H(t)U(t)=i \,\dot U(t)\,,~~~~~~~U(0)=1\;,
	\label{sch-eq-u}
	\xee
must  be preserved under the action of ${\cal U}$. Note that under a
time-dependent quantum canonical transformation the Hamiltonian
undergoes an affine (non-linear) transformation. Hence, unlike the
dynamics, i.e., the Schr\"odinger equation, the energy spectrum is
not preserved.

Next let us study the effect of the transformation induced by (\ref{u}).
In order to compute the transformed Hamiltonian $H'$, one must first
explore the effect of ${\cal U}$ on the position and momentum operators.
A rather lengthy calculation shows that
	\xbea
	x\to x'&:=&{\cal U}\:x\:{\cal U}^\dagger\:=\:{\cal F}_1(x)\;,
	\label{x}\\
	p\to p'&:=&{\cal U}\:p\:{\cal U}^\dagger\:=\:\frac{1}{2}\{{\cal F}_2(x),p\}
	\:=\:\sqrt{{\cal F}_2(x)}\:p\:\sqrt{{\cal F}_2(x)}\;,
	\label{p}
	\xeea
where
	\[ {\cal F}_1(x):=e^{\epsilon(t)f(x)\frac{d}{dx}}\,x\;,~~~~
	{\cal F}_2(x):=f(x)\,e^{\epsilon(t)f(x)\frac{d}{dx}}\,f^{-1}(x)\;.\]
In the derivation of these formulae use is made of
Baker-Campbell-Hausdorff Lemma: 
	\[ e^ABe^A=B+[A,B]+\frac{1}{2!}[A,[A,B]]+\cdots\;,\]
and the identities
	\xbea
	\left[\{f_1(x),p\},f_2(x)\right]&=&-2if_1(x)\frac{d}{dx}f_2(x)\;,\xnn\\
	\left[ \{ f_1(x),p\},\{f_2(x),p\}\right]&=&
	\{ f_3(x),p\}\;,~~~~f_3(x):=-2if_1^2\frac{d}{dx}\left(\frac{f_2}{f_1}\right)\;,
	\xnn
	\xeea
where $f_1$ and $f_2$ are arbitrary functions.

In view of Eqs.~(\ref{trans}), (\ref{x}) and (\ref{p}), one has:
	\xbe
	H'=H'(x',p';t)=H(x',p';t)-\frac{\dot\epsilon(t)}{2}\{f(x),p\}\;.
	\label{h'}
	\xee

Now let us concentrate on a subclass of quantum canonical
transformations of the form (\ref{u}) corresponding to the choice
$f(x)=x$. In this case, Eqs.~(\ref{x}), (\ref{p}), and (\ref{h'})
reduce to
	\xbea
	x\to x'&=&e^{\epsilon(t)}x\;,~~~~~
	p\to p'\:=\:e^{-\epsilon(t)}p\;,
	\label{trans-xp}\\
	H\to H'&=&H(x',p';t)-\frac{\dot\epsilon(t)}{2}\{x,p\}\;.
	\label{trans-h}
	\xeea
For a Hamiltonian of the standard form 
	\xbe
	H=\frac{p^2}{2m}+V(x)\;,
	\label{natural}
	\xee
the latter equation leads to
	\xbe
	H'=\frac{p^2}{2me^{2\epsilon(t)}}+V(e^{\epsilon(t)}x)
	-\frac{\dot\epsilon(t)}{2}\{x,p\}\;.
	\label{H'2}
	\xee
Hence the transformed  Hamiltonian is not of the standard form (\ref{natural}).
It can however be put in this form by the canonical transformation defined
by
	\xbe
	{\cal U'}=\exp\left[\frac{-i}{2}(\dot\epsilon m e^{2\epsilon})x^2\right]\:,
	\label{can2}
	\xee
This leads to
	\xbea
	x\to x''&=&x,~~~~~p\to p''\:=\: p+me^{2\epsilon}\dot\epsilon x\;,
	\label{trans-xp'}\\
	H'\to H''&=&\frac{p^2}{2me^{2\epsilon}}+V(e^{\epsilon}x)+
	\frac{1}{2}\left[ \frac{d}{dt}(me^{2\epsilon}\dot\epsilon)-
	me^{2\epsilon}\dot\epsilon^2\right]\,x^2\;.
	\label{H'3}
	\xeea

In particular for a harmonic oscillator, with $V(x)=m\omega^2x^2/2$,
one has:
	\xbe
	H''=\frac{p^2}{2me^{2\epsilon}}+
	\frac{1}{2}\left[\frac{d}{dt}(me^{2\epsilon}\dot\epsilon)
	+me^{2\epsilon}(\omega^2-\dot\epsilon^2)\right]x^2\;.
	\label{osc}
	\xee
Note that here the mass $m$ and frequency $\omega$ may depend
on time, i.e., $m=m(t),~\omega=\omega(t)$.

Next let us choose $\epsilon(t)=\ln[m_0-m(t)]/2$, for a positive constant
$m_0$, so that $me^{2\epsilon}=m_0$. This choice leads to a harmonic
oscillator with constant mass $m_0$ and a frequency:
	\xbe
	\Omega:=\sqrt{\ddot\epsilon-\dot\epsilon^2+\omega^2}\:,
	\label{Omega}
	\xee
namely one has:
	\xbe
	H''=\frac{p^2}{2m_0}+\frac{1}{2}\,m_0\Omega^2x^2\;.
	\label{H'4}
	\xee
Requiring $\Omega$ to be independent of time, one can exactly solve
the Schr\"odinger equation for $H''$ which is now time-independent.
Transforming back the solution of the Schr\"odinger equation
for $H''$ using the canonical transformation ${\cal U}'':=
({\cal U}'{\cal U})^\dagger$ and Eq.~(\ref{trans-u}), one then finds
the exact solution of  the Schr\"odinger equation for the original
harmonic oscillator. Thus the requirement $\Omega=\Omega_0=$ const.\
corresponds to a class of exactly solvable time-dependent harmonic
oscillators. The solution of the Schr\"odinger equation for time-dependent
harmonic oscillator has been the subject of research since 1940's,
\cite{c-k,osc,abd}. But  for arbitrary choices of mass $m(t)$ and frequency
$\omega(t)$ a closed expression for the time-evolution operator
is not yet known.

Next let us re-express the condition $\Omega=\Omega_0$ in terms of
$m$ and $\omega$. This leads to
	\xbe
	\ddot\epsilon-\dot\epsilon^2+\alpha^2=0\;,~~~~\alpha:=
	\sqrt{\omega^2-\Omega_0^2}\;,
	\label{condi}
	\xee
or alternatively
	\xbe
	\omega=\sqrt{\Omega_0^2+\frac{\ddot m}{2m}-
	\left(\frac{\dot m}{2m}\right)^2}\;.
	\label{condi-1}
	\xee
Hence, according to the above argument, the  time-dependent oscillators
whose mass $m$ and frequency $\omega$ satisfy (\ref{condi-1}) are
canonically equivalent to the time-independent
harmonic oscillator (\ref{H'4}) with $\Omega=\Omega_0$.

Next let us consider the case where the frequency $\omega$ is constant.
Then Eq.~(\ref{condi-1}) can be easily integrated to yield
	\xbe
	m(t)=m_0\left(\mu e^{\alpha t}+\nu e^{-\alpha t}\right)^2\;,
	\label{m=}
	\xee
where $\mu$ and $\nu$ are constants. Clearly, the Caldirola-Kanai
oscillator \cite{c-k} whose mass depends exponentially on time, i.e.,
$m=m_0e^{\gamma t}$ belongs to this class of oscillators. In fact,
Colegrave and Abdalla have considered using the canonical
transformation (\ref{trans-xp}) to treat the oscillators with time-dependent
mass and fixed frequency, and in particular the Caldirola-Kanai oscillator,
\cite{abd}. However, they perform the canonical transformation within the
classical context and then quantize the Hamiltonian. Fortunately the Hermiticity
requirement determines the quantum  Hamiltonian uniquely. Hence the lack of
knowledge about the precise unitary transformation corresponding to this
canonical transformation does not play much of a role in their analysis.

It should be emphasized that  the choice $f(x)=x$ only determines a
small class of  quantum canonical transformations of the form (\ref{u}).
As it is seen from Eqs.~(\ref{x}) and (\ref{p}) the transformations
induced on the position and momentum operators depend in a complicated
manner on $f(x)$. Some other choices of $f(x)$ for which these transformations
can be calculated in a closed form are
	\xbe
	\begin{array}{cccc}
	f(x)=x^2 :&\left\{\begin{array}{cc}
	x&\rightarrow x'=\frac{x}{1-\epsilon(t) x}\\
	p&\rightarrow p'=[1-\epsilon(t)x]\,p\,[1-\epsilon(t)x]
	\end{array}\right\}&{\rm for}&\epsilon(t)x<1\\
	&&&\\
	f(x)=e^{-\lambda(t) x}:&\left\{\begin{array}{cc}
	x&\rightarrow x'=\frac{1}{\lambda(t)}\ln\left[e^{\lambda(t) x}+
	\epsilon(t)\lambda(t)\right]\\
	p&\rightarrow p'=\sqrt{1+\frac{\epsilon(t)e^{\lambda(t)x}}{\lambda(t)}}
	\,p\,\sqrt{1+\frac{\epsilon(t)e^{\lambda(t)x}}{\lambda(t)}}
	\end{array}\right\}&{\rm for}&\epsilon(t)\lambda(t)<1
	\end{array}\;,
	\label{eg}
	\xee
where $\lambda$ is a positive real-valued function of  time. As seen from
these formulae, the effect of these canonical transformations on the kinetic
part $p^2/2m$ of the Hamiltonian is to make the mass $m$ to also depend
on the position. This is precisely what happens when one considers a free particle
moving on a line with a nontrivial metric $g$. In this case the quantum
Hamiltonian is given by \cite{super}
	\xbe
	{\cal H}= \frac{1}{2m}\left[ g^{-1/4}p\,g^{-1/2}p\,g^{-1/4}\right]\;.
	\label{curved}
	\xee
It is uniquely determined by the classical Hamiltonian ${\cal H}_c=p^2/(2mg)$
and the self-adjointness requirement with respect to the measure $\sqrt{g}dx$.

In view of  Eqs.~(\ref{curved}) and~(\ref{p}), one can easily infer the fact that
under the canonical transformations (\ref{u}) a free particle in $\xR$ with
the trivial metric ($g(x)=1$) is mapped to a free particle in $\xR$ with a
metric $g=[{\cal F}_2]^{-2}$. The converse of this statement is
also true in the sense that for an arbitrary metric $g=g(x;t)$, there is a canonical
transformation of the form (\ref{u}) which maps the problem to the ordinary
one-dimensional free particle problem provided that one can solve the
pseudo-differential equation
	\[f(x)\,e^{\epsilon(t)f(x)\frac{d}{dx}}\,f^{-1}(x)={\cal F}_2(x)=
	[g({\cal F}_1(x);t)]^{-1/2}=\left[g(e^{\epsilon(t)f(x)\frac{d}{dx}}\,x;t)
	\right]^{-1/2}\]
for $f(x)$.
For the examples listed in (\ref{eg}), one has
	\xbea
	f=x^2~~~~   &\Longleftrightarrow& g=[1-\epsilon(t)x]^{-4}\;,\xnn\\
	f=e^{-\lambda(t)x} &\Longleftrightarrow& g=\left[
	1+\frac{\epsilon(t)e^{\lambda(t)x}}{\lambda(t)}\right]^{-2}\;.
	\xnn
	\xeea

The  quantum canonical transformations of the form (\ref{u}) play an
important role in finding exact solutions of the Schr\"odinger equation for the
class of time-dependent harmonic oscillators defined by (\ref{condi-1}). 
They may also be used to show the canonical equivalence of the
one-dimensional quantum mechanics of a free particle with position
(and time) dependent mass with that of a free particle with constant mass.
The direct generalization of the analysis presented in this article to higher
dimensions should not be difficult.  In particular, one might study the
implications of the results for the quantum dynamics of a free particle
moving on a manifold with a nontrivial metric\footnote{Note that here
a nontrivial metric means $g_{ij}\neq\delta_{ij}$.}.

\section*{Acknowledgements}
I wish to thank Dr.~M.~Razavi for interesting discussions and acknowledge
the financial support of the Killam Foundation of Canada.


\end{document}